\documentclass[thmsa,12pt]{article}
\usepackage{amssymb}

\usepackage{amsmath}

\input{tcilatex}

\begin{document}

\setcounter{page}{0} \topmargin0pt \oddsidemargin5mm \renewcommand{%
\thefootnote}{\fnsymbol{footnote}} \newpage \setcounter{page}{0} 
\begin{titlepage}
\begin{flushright}
YITP-SB-01-15 \\
\end{flushright}
\vspace{0.5cm}
\begin{center}
{\Large {\bf Loop symmetry of integrable vertex models\\
at roots of unity} }

\vspace{0.8cm}
{ \large Christian Korff and Barry M. McCoy}

\vspace{0.5cm}
{\em C.N. Yang Institute for Theoretical Physics\\
State University of New York at Stony Brook\\
Stony Brook, N.Y. 11794-3840}
\end{center}
\vspace{0.2cm}

\renewcommand{\thefootnote}{\arabic{footnote}}
\setcounter{footnote}{0}

\begin{abstract}
It has been recently discovered in the context of the six vertex or XXZ model in the fundamental
representation that new symmetries arise when the anisotropy parameter $(q+q^{-1})/2$ is
evaluated at roots of unity $q^{N}=1$. These new symmetries have been linked to an
$U(A^{(1)}_1)$ invariance of the transfer matrix and the corresponding spin-chain Hamiltonian.
In this paper these results are generalized for odd primitive roots of unity  to all vertex models 
associated with trigonometric solutions of the Yang-Baxter equation by invoking representation
 independent methods which only take the algebraic structure of the underlying quantum
 groups $U_q(\hat g)$ into account.
Here $\hat g$ is an arbitrary Kac-Moody algebra. Employing the notion of the boost
operator it is then found that the Hamiltonian and the transfer matrix of the integrable model 
are invariant under the action of $U(\hat{g})$. For the simplest case $\hat g=A_1^{(1)}$ the 
discussion is also extended to even primitive roots of unity.
\par\noindent
\end{abstract}
\vfill{ \hspace*{-9mm}
\begin{tabular}{l}
\rule{6 cm}{0.05 mm}\\
korff@insti.physics.sunysb.edu \\
mccoy@insti.physics.sunysb.edu
\end{tabular}}
\end{titlepage}
\newpage

\section{Introduction}

The six-vertex \cite{six} or XXZ model \cite{XXZ} with periodic boundary
conditions as defined by the following spin-chain Hamiltonian 
\begin{equation}
\mathcal{H}_{\text{XXZ}}^{s=1/2}=\sum_{j=1}^{L}\left\{ \sigma _{j}^{x}\sigma
_{j+1}^{x}+\sigma _{j}^{y}\sigma _{j+1}^{y}+\frac{q+q^{-1}}{2}\left( \sigma
_{j}^{z}\sigma _{j+1}^{z}-1\right) \right\} \;,\quad L+1\equiv 1  \label{xxz}
\end{equation}
has been subject to extensive studies for a long time. Here $\sigma
_{j}^{x},\sigma _{j}^{y},\sigma _{j}^{z}$ are the Pauli matrices acting on
the $j^{\text{th}}$ lattice site. Surprisingly the model and its underlying
symmetries are still not fully understood. Baxter already noted in 1973 \cite
{Baxter2}\ that the model besides its integrable structure for generic
anisotropy parameter $(q+q^{-1})/2$ shows additional symmetries when $q$
becomes an $N^{\text{th}}$ primitive root of unity (i.e. $N$ is the smallest
integer such that $q^{N}=1$). Despite numerous articles addressing the
energy spectrum and the problem of completeness of the eigenstates for
generic $q,$ the symmetry governing the root of unity case has just recently
been discovered in \cite{DFM,FM}. The key results obtained by algebraic and
numerical methods in the latter articles are the following,

\begin{enumerate}
\item  As $q$ approaches a root of unity the transfer matrix of the
six-vertex model as well as the associated Hamiltonian exhibit an $%
U(A_{1}^{(1)})$ invariance at level zero. (It is for this reason that we
have mentioned loop symmetry instead of affine symmetry in the title.) For
total spin values being a multiple of\footnote{%
Note that in the articles \cite{DFM,FM} a different convention to
parametrize the roots of unity had been chosen. The power $N$ in the latter
articles correponds to the power $N^{\prime }$ in this work.} 
\begin{equation}
N^{\prime }:=\left\{ 
\begin{array}{cc}
N/2\;, & N\text{ even} \\ 
N\;, & N\text{ odd}
\end{array}
\right.  \label{Np}
\end{equation}
the generators of this symmetry algebra can be constructed from the quantum
group $U_{q}(A_{1}^{(1)})$ associated with the $R$-matrix of the six-vertex
model as $q^{N}\rightarrow 1$. In addition, the symmetry algebra preserves
the momentum, i.e. the $U(A_{1}^{(1)})$ generators commute with the shift
operator.

\item  In the framework of the Bethe Ansatz \cite{Bethe} the degeneracies
manifest themselves in additional string solutions possessing zero energy,
which are called exact complete $N^{\prime }$-strings and were first found
by Baxter \cite{Baxter2} (see also the review of Takahashi \cite{Taka}).
However, the link between these string solutions and the above symmetry
algebra had not been recognized. Moreover, the exact complete $N^{\prime }$%
-strings lead to a simultaneous vanishing of the numerator and the
denominator inside Bethe's equation, the latter therefore fails to determine
the complete set of eigenstates. It has been demonstrated in \cite{FM} how
additional equations can be derived from Bethe's equation in the limit $%
q^{N}\rightarrow 1$ which then allow the determination of the real parts of
the exact complete $N^{\prime }$-strings.
\end{enumerate}

\noindent All of the above observations have been made in the context of the
fundamental representation, i.e. for spin $s=1/2$, and some of the proofs in 
\cite{DFM} take explicitly advantage of features only present in this
particular case. However, one might expect that the degeneracies are of a
more general nature. For example the spin $s=1$ XXZ model first introduced
by Fateev and Zamolodchikov \cite{ZF} is closely related at roots of unity $%
q=\exp (i\pi /3)$ to the 3-state super-integrable chiral Potts model for
which similar degeneracies have been observed numerically in \cite{DKM}. The
Hamiltonian in this case looks considerably more complicated than the one in
the fundamental representation \cite{ZF}, 
\begin{eqnarray}
\mathcal{H}_{\text{XXZ}}^{s=1} &=&\sum_{j=1}^{L}\left\{ \mathbf{S}_{j}%
\mathbf{S}_{j+1}-(\mathbf{S}_{j}\mathbf{S}_{j+1})^{2}-2(1-\Delta ^{2})\left[
S_{j}^{z}S_{j+1}^{z}-(S_{j}^{z}S_{j+1}^{z})^{2}+2(S_{j}^{z})^{2}\right]
\right.  \notag \\
&&\left. +2(1+\Delta )\left[ \left(
S_{j}^{x}S_{j+1}^{x}+S_{j}^{y}S_{j+1}^{y}\right)
S_{j}^{z}S_{j+1}^{z}+S_{j}^{z}S_{j+1}^{z}\left(
S_{j}^{x}S_{j+1}^{x}+S_{j}^{y}S_{j+1}^{y}\right) \right] \right\}
\label{xxz2}
\end{eqnarray}
Here $\mathbf{S}_{j}=(S_{j}^{x},S_{j}^{y},S_{j}^{z})$ is the spin operator
in the vector representation $s=1$ at site $j$. The XXZ model for arbitrary
spin has been investigated in \cite{KiResh,FYF}. The corresponding
Hamiltonians for spin $s>1$ have not been written down in terms of spin
operators but can be defined through the transfer matrix of the associated
statistical model. Also here the problem of the degeneracies and the
underlying symmetry at roots of unity has not been addressed. Besides the
extension to arbitrary spin one can also consider the case of higher rank.
For the fundamental representation $V_{\lambda _{1}}=\mathbb{C}^{n+1}$ the
R-matrix associated with the $A_{n}^{(1)}\equiv \widehat{sl}_{n+1}$ vertex
models has been found in \cite{slnR} and the corresponding spin-chain
Hamiltonian in \cite{Perk} , 
\begin{eqnarray}
\mathcal{H}_{A_{n}^{(1)}} &=&\sum_{j=1}^{L}\,\left\{ \tsum_{k\neq l}\left[ 
\mathbb{E}_{j}^{kl}\otimes \mathbb{E}_{j+1}^{kl}+\mathbb{E}_{j}^{kl}\otimes 
\mathbb{E}_{j+1}^{lk}+i\mathbb{E}_{j}^{kl}\otimes i\mathbb{E}_{j+1}^{kl}-i%
\mathbb{E}_{j}^{kl}\otimes i\mathbb{E}_{j+1}^{lk}\right] \right.  \notag \\
&&\left. +\frac{q+q^{-1}}{2}\left( \tsum_{k<l}\left[ \mathbb{E}%
_{j}^{kk}\otimes \mathbb{E}_{j+1}^{ll}-\mathbb{E}_{j}^{ll}\otimes \mathbb{E}%
_{j+1}^{kk}\right] -\tsum_{k}\mathbb{E}_{j}^{kk}\otimes \mathbb{E}%
_{j+1}^{kk}\right) \right\}  \label{sln}
\end{eqnarray}
where $\mathbb{E}^{kl}$, $1\leq k,l\leq n+1$ denote the $(n+1)\times (n+1)$
unit matrices, whose entries are all zero except for the entry in the $k^{%
\text{th}}$ row and $l^{\text{th}}$ column which is equal to one. For $n=1$
one recovers the XXZ Hamiltonian (\ref{xxz}). Many R-matrices belonging to
algebras different from $\hat{g}=A_{n}^{(1)}$ have been investigated in e.g. 
\cite{Smat}.\bigskip

\noindent In this article we demonstrate that the above observations are
indeed of a very general nature and not only can be extended for the XXZ
model to arbitrary spin but also to the much wider class of integrable
vertex models associated with the quantum groups $U_{q}(\hat{g})$, where $%
\hat{g}$ is an arbitrary Kac-Moody algebra \cite{Kac}. The relation between
integrable models and the quantum groups $U_{q}({\hat{g})}$ was first
established by Drinfel'd \cite{Drin} and Jimbo \cite{Jimbo} who studied
trigonometric solutions of the Yang-Baxter equation \cite{YB}, 
\begin{equation}
R_{12}(u)R_{13}(u+v)R_{23}(v)=R_{23}(v)R_{13}(u+v)R_{12}(u)\;.  \label{YB}
\end{equation}
The Yang-Baxter equation is an operator identity over $V_{1}\otimes
V_{2}\otimes V_{3}$ with\ $R_{ab}(u)$ acting on $V_{a}\otimes V_{b}$ and $%
V_{a}\cong V$ being some representation space of $U_{q}({\hat{g})}$. An
integrable $L^{\prime }\times L$ vertex model is now implicitly defined when
interpreting the matrix elements of the solution $R(u)$ as Boltzmann weights
and taking the partition function and transfer matrix to be 
\begin{equation}
Z=\limfunc{Tr}\nolimits_{V^{\otimes L}}T^{L^{\prime }}(u),\quad \quad T(u):=%
\limfunc{Tr}\nolimits_{V_{0}}R_{0L}(u)R_{0L-1}(u)\cdots R_{01}(u)\;.
\label{trans}
\end{equation}
The transfer matrix acts on the tensor product space $V^{\otimes L}\equiv
V_{1}\otimes V_{2}\cdots \otimes V_{L}$ and the trace is taken over the
boundary values encoded in the auxiliary space $V_{0}$. As is well known (%
\ref{YB}) ensures that the transfer matrix when evaluated at different
spectral parameters $u$ commutes with itself rendering the model (\ref{trans}%
) integrable \cite{Baxter,QISM}. The corresponding 'spin'-chain Hamiltonian
is now generically given by 
\begin{equation}
\mathcal{H}=i\left. \frac{\partial }{\partial u}\ln T(u)\right| _{u=0}+\text{%
const.}  \label{Ham}
\end{equation}
Up to possible scaling factors depending on different conventions and an
additive constant depending on the normalization of the R-matrix this
definition specializes for $\hat{g}=A_{n}^{(1)}$ to the stated examples (\ref
{xxz}), (\ref{xxz2}) and (\ref{sln}).

It is important to note that in the correspondence between quantum groups
and integrable vertex models the quantum group $U_{q}({\hat{g}})$ does not
define a symmetry of the model and that its generators do not commute with
either the Hamiltonian or the transfer matrix. This lack of symmetry comes
ultimately from the fact that by construction the transfer matrix (\ref
{trans}) and Hamiltonian (\ref{Ham}) we are considering are translational
invariant in the appropriate space $V^{\otimes L}$ whereas for generic $q$
the quantum group $U_{q}({\hat{g}})$ does not act in a translational
invariant fashion on this space. The situation changes considerably when the
deformation parameter $q$ approaches a primitive root of unity, $q^{N}=1$
with $N\geq 3$ being odd. Then symmetry generators can be extracted from $%
U_{q}(\hat{g})$ in the limiting process $q^{N}\rightarrow 1$ which are
translation invariant and which generate the algebra $U({\hat{g}})$ at level
zero. \bigskip

\noindent The paper is organized as follows. In Section 2 we construct the
symmetry algebra $U(\hat{g})$ for odd roots of unity and highest weight
representations $\lambda $ obeying $\lambda (h_{i})\equiv 0\func{mod}N$ with 
$\lambda (h_{i})$ being the eigenvalues of the Cartan subalgebra generators $%
h_{i}$. In Section 3 we prove the translational invariance of the symmetry
generators when they act on the space $V^{\otimes L}$. In Section 4 we
demonstrate for any algebra $\hat{g}$ in a completely generic and
representation independent way that the transfer matrix and Hamiltonian
associated with $U_{q}({\hat{g}})$ commute with the generators of $U({\hat{g}%
})$ when $q$ is a root of unity. This proof makes use of the boost operator
(e.g. \cite{boost}) and the quantum group theoretical structure underlying
the Yang-Baxter equation as developed by Drinfel'd \cite{Drin} and Jimbo 
\cite{Jimbo}. Some results of their construction which are relevant to our
discussion are reviewed in the appendix. For the simplest case $\hat{g}%
=A_{1}^{(1)}$, i.e. the XXZ model, we discuss for arbitrary spin how the
results can also be extended to even roots of unity. Finally we conclude in
Section 5 with a discussion of our results. By making contact with the
representation theory of the symmetry algebra we argue that for all
untwisted algebras the degeneracies of the energy eigenstates should be
given by powers of the dimension of the fundamental representation $\left(
\dim \lambda \right) ^{l}$ where $l$ is some integer depending on the
multiplet of the energy eigenstate.

\section{Constructing $U(\hat{g})$ from $U_{q}(\hat{g})$ at roots of unity}

We begin by reviewing the basic definition of $U_{q}(\hat{g})$ for arbitrary
Kac-Moody algebras $\hat{g}$ in order to introduce our notation (further
details can be found in the original references \cite{Drin,Jimbo} and in
numerous monographs e.g. \cite{CP,EFK}). The quantum universal enveloping
algebra $U_{q}(\hat{g})$ is the algebra of power series in the
Chevalley-Serre generators $\{e_{i},f_{i},h_{i}\}_{i=0}^{\limfunc{rank}\hat{g%
}-1}\cup \{1\}$ subject to the following commutation relations:

\begin{itemize}
\item[(Q1)]  Let $A$ denote the Cartan matrix associated with the Kac-Moody
algebra $\hat{g}$. Then 
\begin{equation*}
\lbrack h_{i},h_{j}]=0\,,\quad \lbrack h_{i},e_{j}]=A_{ij}e_{j}\,,\quad 
\text{and\quad }[h_{i},e_{j}]=-A_{ij}f_{j}\;.
\end{equation*}

\item[(Q2)]  Considering only highest weight representations in which the $%
h_{i}$'s act as multiplication operators, we introduce the exponentiated
generators $q_{i}^{h_{i}}$ with $q_{i}=q^{\alpha _{i}^{2}/2}$ and $\alpha
_{i}$ denoting a simple root of $\hat{g}$. Then one requires 
\begin{equation*}
\lbrack e_{i},f_{j}]=\delta _{ij}\,\dfrac{q_{i}^{h_{i}}-q_{i}^{-h_{i}}}{%
q_{i}^{1}-q_{i}^{-1}}\;.
\end{equation*}
For simplicity we choose throughout this paper the normalization convention $%
\alpha ^{2}=2$ for \emph{short }roots. That is, for long roots $q_{i}$ might
equal the powers $q,q^{2},q^{3}$ in the deformation parameter.

\item[(Q3)]  In addition, the generators ought to satisfy the quantum
Chevalley-Serre relations 
\begin{eqnarray*}
\sum_{n=0}^{1-A_{ij}}(-1)^{n}\QATOPD[ ] {1-A_{ij}}{n}_{q_{i}}\,e_{i}^{n}%
\,e_{j}\,e_{i}^{1-A_{ji}-n} &=&0\;,\quad i\neq j \\
\sum_{n=0}^{1-A_{ij}}(-1)^{n}\QATOPD[ ] {1-A_{ij}}{n}_{q_{i}}\,f_{i}^{n}%
\,f_{j}\,f_{i}^{1-A_{ij}-n} &=&0\;,\quad i\neq j
\end{eqnarray*}
Here $q$-integers have been introduced, 
\begin{equation}
\QATOPD[ ] {m}{n}_{q}:=\frac{[m]_{q}!}{\left[ n\right] _{q}!\left[ m-n\right]
_{q}!}\;,\quad \lbrack n]_{q}!:=\prod_{k=1}^{n}[k]_{q}\;,\quad \lbrack
n]_{q}:=\frac{q^{n}-q^{-n}}{q^{1}-q^{-1}}\;.
\end{equation}
\end{itemize}

We will now focus on the case where the deformation parameter $q$ takes the
value of a primitive root of unity, $q^{N}=1$. It is known that in this case
the elements $e_{i}^{N^{\prime }}$ and $f_{i}^{N^{\prime }}$ are central
elements. We are here interested in representations in which these central
elements may be set equal to zero and for these representations it has been
shown \cite{Lus}\ that the generators 
\begin{equation}
e_{i}^{(N^{\prime })}:=\frac{e_{i}^{N^{\prime }}}{[N^{\prime }]_{q_{i}}!}%
\;,\quad f_{i}^{(N^{\prime })}:=\frac{f_{i}^{N^{\prime }}}{[N^{\prime
}]_{q_{i}}!}\;,\quad \text{and\quad }h_{i}/N^{\prime }  \label{alg}
\end{equation}
stay well defined in the limit $q^{N}\rightarrow 1$. As it was first
observed in \cite{DFM} for $\hat{g}=A_{1}^{(1)}$ in the fundamental
representation the above set (complemented by unity) generates the
non-deformed enveloping algebra $U(\hat{g})$ provided one restricts oneself
to highest weight representations $\left| \lambda \right\rangle $ satisfying%
\footnote{%
For the case $\hat{g}=A_{1}^{(1)}$ considered in \cite{DFM}\ this
corresponds to condition $S^{z}=0\func{mod}N^{\prime }$ with $%
h_{1}=-h_{0}=2S^{z}$ and $N$ even.} 
\begin{equation}
\lambda (h_{i})=0\func{mod}N\;.  \label{con}
\end{equation}
We impose this condition since we will ultimately make use of $%
q_{i}^{h_{i}}=1$ in the calculations (see for example equation (\ref{Pinv})
in the next section). Note also that this condition will ultimately have to
hold for tensor products of highest weight representations, since we are
going to consider the action of $U(\hat{g})$ on $V^{\otimes L},$ see the
next section. The proof follows along the lines of \cite{DFM} and we recall
here the key steps in order to keep this article self-contained. First we
investigate the commutation relations between the generators $%
e_{i}^{(N^{\prime })},f_{i}^{(N^{\prime })}$ starting from the following
relation valid for all $q$ \cite{CK}, 
\begin{equation}
\lbrack e_{i}^{(m)},f_{i}^{(n)}]=\sum_{l=1}^{\min
(m,n)}\,f_{i}^{(n-l)}\,e_{i}^{(m-l)}\prod_{r=1}^{l}\frac{%
q_{i}^{h_{i}}q_{i}^{m-n-r+1}-q_{i}^{-h_{i}}q_{i}^{-m+n+r-1}}{%
q_{i}^{r}-q_{i}^{-r}}\;.
\end{equation}
Choosing a highest weight such that (\ref{con}) is satisfied one obtains in
the limit $q^{N}\rightarrow 1,$ 
\begin{equation}
\lim_{q^{N}\rightarrow 1}[e_{i}^{(N^{\prime })},f_{i}^{(N^{\prime
})}]=(-1)^{N^{\prime }-1}\lim_{q^{N}\rightarrow 1}\frac{%
q_{i}^{h_{i}}-q_{i}^{-h_{i}}}{q_{i}^{N^{\prime }}-q_{i}^{-N^{\prime }}}%
=(-1)^{N^{\prime }-1}q_{i}^{h_{i}}\frac{h_{i}}{N^{\prime }}\;.  \label{comm}
\end{equation}
Furthermore, one proves easily from (Q1) by induction that 
\begin{equation}
\lbrack h_{i},e_{j}^{(N^{\prime })}]=N^{\prime }\,A_{ij}e_{j}^{(N^{\prime
})}\quad \text{and\quad }[h_{i},f_{j}^{(N^{\prime })}]=-N^{\prime
}\,A_{ij}f_{j}^{(N^{\prime })}\;.
\end{equation}
It remains to verify the Chevalley-Serre relations for $U(\hat{g})$. For
this purpose we employ Lustzig's higher order Chevalley-Serre relations \cite
{Lus} which are valid for generic $q$. Let $m>-A_{ij}n,\;n\geq 1$ then the
generators $e_{i}^{(n)}:=e_{i}^{n}/[n]_{q_{i}}!$ satisfy 
\begin{equation}
e_{i}^{(m)}e_{j}^{(n)}=\sum_{k=0}^{-nA_{ij}}C_{m-k}(q_{i})%
\,e_{i}^{(k)}e_{j}^{(n)}e_{i}^{(m-k)},  \label{Lus}
\end{equation}
where the coefficient function is given by 
\begin{equation}
C_{s}(q)=\sum_{l=0}^{m+A_{ij}n-1}(-1)^{s+l+1}q^{-s(l+1-A_{ij}n-m)+l}\QATOPD[ 
] {s}{l}_{q}
\end{equation}
Choosing $m=1-A_{ij},n=1$ one recovers the usual quantum Chevalley-Serre
relations (Q3). Suppose now that the indeterminate $q$ approaches a root of
unity $q^{N}\rightarrow 1$. Setting $m=N^{\prime }(1-A_{ij})$ and $%
n=N^{\prime }$ one verifies for the coefficient function, 
\begin{equation}
\lim_{q^{N}\rightarrow 1}C_{s}(q_{i})=\left\{ 
\begin{array}{cc}
(-1)^{s+1}q_{i}^{s(N^{\prime }-1)}\;,\, & \text{if\ }s=0\func{mod}N^{\prime }
\\ 
0\;, & \text{else}
\end{array}
\right. \;.  \label{A}
\end{equation}
We now rewrite the higher order Chevalley-Serre equations in terms of powers
of the operators $e_{i}^{(N^{\prime })}$ by employing the identity 
\begin{equation}
e_{i}^{(N^{\prime }s)}=\frac{[N^{\prime }]_{q_{i}}!^{s}}{[N^{\prime
}s]_{q_{i}}!}e_{i}^{(N^{\prime })s}\text{\quad with\quad }%
\lim_{q^{N}\rightarrow 1}\frac{[N^{\prime }]_{q_{i}}!^{s}}{[N^{\prime
}s]_{q_{i}}!}=\frac{q_{i}^{N^{\prime 2}\frac{s(s-1)}{2}}}{s!}\;.  \label{B}
\end{equation}
Plugging the results (\ref{A}) and (\ref{B}) into equation (\ref{Lus}) one
derives the desired Chevalley-Serre relations of the non-deformed enveloping
algebra $U(\hat{g})$ up to certain sign factors, 
\begin{eqnarray}
e_{i}^{(N^{\prime })(1-A_{ij})}e_{j}^{(N^{\prime })}
&=&\sum_{n=0}^{-A_{ij}}(-1)^{N^{\prime
}(1-A_{ij}-n)+1}q_{i}^{(1-A_{ij}-n)N^{\prime }(N^{\prime }-1)}  \notag \\
&&\quad \times q_{i}^{-nN^{\prime 2}(1-A_{ij}-n)}\binom{1-A_{ij}}{n}%
e_{i}^{(N^{\prime })n}e_{j}^{(N^{\prime })}e_{i}^{(N^{\prime
})(1-A_{ij}-n)}\;.  \label{chevs}
\end{eqnarray}
An analogous equation holds for the generators $f_{i}^{(N^{\prime })}$. In
order to make contact to the Chevalley-Serre relations of $U(\hat{g})$ one
has to discuss carefully the cancellation of the minus signs in the r.h.s.
of the above equation. We distinguish the following three cases.

\begin{description}
\item[$N$ odd. ]  For odd roots of unity one recovers the correct sign $%
(-1)^{(1-A_{ij}-n)+1}$ needed for the Chevalley-Serre relations in the
r.h.s. of (\ref{chevs}). For this case we can therefore conclude that the
algebra spanned by the elements (\ref{alg}) can be identified with the
non-deformed enveloping algebra $U(\hat{g})$ for all Kac-Moody algebras.

\item[$N^{\prime }$ even. ]  For even roots of unity one has $q^{N^{\prime
}}=-1$. Provided that $N^{\prime }$ even and $q_{i}^{N^{\prime }}=-1$ for
all $i$ one obtains the correct sign factor also in this case. The latter
condition requires $\hat{g}$ to be either simply-laced, i.e. $\hat{g}%
=A_{n}^{(1)},D_{n}^{(1)},E_{6,7,8}^{(1)},A_{2n}^{(2)}$, or to be one of the
non simply-laced algebras $\hat{g}=G_{2}^{(1)},D_{4}^{(3)}$.

\item[$N^{\prime }$ odd. ]  In the remaining case of even roots of unity
with $N^{\prime }$ odd one obtains the sign factor $%
(-1)^{(n+1)(1-A_{ij}-n)+1}$. In general this will not reproduce the correct
Chevalley-Serre relations. For the simplest case $\hat{g}=A_{1}^{(1)},$
however, the signs work out correctly which can be explicitly checked by
using $A_{ij}=-2$ for $i\neq j,$ compare also \cite{DFM}. But now one has to
pay attention to the sign in (\ref{comm}).
\end{description}

\section{Translation invariance}

We now establish that the action of the constructed symmetry algebra $U(\hat{%
g})$ is translation invariant. That is, given some representation space $V$
of $U_{q}(\hat{g})$ we consider its $L$-fold tensor product $V^{\otimes L}$
and then show in the limit $q^{N}\rightarrow 1$ that the action of the
symmetry algebra on this space commutes with the shift-operator defined as 
\begin{equation}
\Pi :V_{1}\otimes V_{2}\cdots \otimes V_{L}\rightarrow V_{2}\otimes \cdots
V_{L}\otimes V_{1}\;.  \label{shift}
\end{equation}
As a preliminary step we first recall the action of $U_{q}(\hat{g})$ on $%
V^{\otimes L}$. The latter is determined by the fact that the quantum groups
defined through (Q1)-(Q3) are endowed with the structure of an Hopf algebra 
\cite{Drin,Jimbo}. This requires in general the notion of a co-unit $\bar{e}$%
, an anti-pode $\gamma $ and a coproduct $\Delta $. We will only need the
concept of the latter which establishes an algebra homomorphism $U_{q}(\hat{g%
})\rightarrow U_{q}(\hat{g})\otimes U_{q}(\hat{g})$. There are different
conventions in the literature how to define the coproduct and we choose the
one most convenient for our purposes, 
\begin{eqnarray}
\Delta (h_{i}) &=&h_{i}\otimes 1+1\otimes h_{i}\quad  \notag \\
\Delta (e_{i}) &=&e_{i}\otimes q_{i}^{-\frac{h_{i}}{2}}+q_{i}^{\frac{h_{i}}{2%
}}\otimes e_{i}\quad \text{and\quad }\Delta (f_{i})=f_{i}\otimes q_{i}^{-%
\frac{h_{i}}{2}}+q_{i}^{\frac{h_{i}}{2}}\otimes f_{i}  \notag \\
\Delta (\,1\,) &=&1\otimes 1  \label{copo}
\end{eqnarray}
Following \cite{Jimbo} we can now use the coproduct $\Delta $ iteratively to
generate higher tensor products by setting 
\begin{equation}
\Delta ^{(L)}=(\Delta \otimes \mathbf{1}_{L-2})\Delta ^{(L-1)}\quad \text{%
with\quad }\Delta ^{(2)}\equiv \Delta \;.  \label{copL}
\end{equation}
In fact, this defines an algebra homomorphism $\Delta ^{(L)}:U_{q}(\hat{g}%
)\rightarrow U_{q}(\hat{g})^{\otimes L}$ and the generators acting on $%
V^{\otimes L}$ then explicitly read, 
\begin{eqnarray}
\Delta ^{(L)}(e_{i}) &\equiv &E_{i}=\sum_{n=1}^{L}E_{i}(n)\;,\quad
E_{i}(n):=q_{i}^{\frac{h_{i}}{2}}\otimes \cdots q_{i}^{\frac{h_{i}}{2}%
}\otimes \underset{n^{\text{th}}}{e_{i}}\otimes q_{i}^{-\frac{h_{i}}{2}%
}\cdots \otimes q_{i}^{-\frac{h_{i}}{2}}  \notag \\
\Delta ^{(L)}(f_{i}) &\equiv &F_{i}=\sum_{n=1}^{L}F_{i}(n)\;,\quad
F_{i}(n):=q_{i}^{\frac{h_{i}}{2}}\otimes \cdots q_{i}^{\frac{h_{i}}{2}%
}\otimes \underset{n^{\text{th}}}{f_{i}}\otimes q_{i}^{-\frac{h_{i}}{2}%
}\cdots \otimes q_{i}^{-\frac{h_{i}}{2}}  \notag \\
\Delta ^{(L)}(q_{i}^{h_{i}}) &\equiv
&q_{i}^{H_{i}}=\prod_{n=1}^{L}q_{i}^{H_{i}(n)}\;,\quad
q_{i}^{H_{i}(n)}:=1\otimes \cdots 1\otimes \underset{n^{\text{th}}}{%
q_{i}^{h_{i}}}\otimes 1\cdots \otimes 1\;.  \label{gen}
\end{eqnarray}
For completeness we state also the explicit form of the symmetry generators (%
\ref{alg}) for the $L$-fold tensor product. Starting from the following
relation which is easily proved by induction, 
\begin{equation}
\Delta (e_{i}^{n})=\sum_{k=0}^{n}\QTATOPD[ ] {n}{k}_{q_{i}}%
\,e_{i}^{k}q_{i}^{(n-k)\frac{h_{i}}{2}}\otimes e_{i}^{n-k}q_{i}^{-k\frac{%
h_{i}}{2}}=\sum_{k=0}^{n}\QTATOPD[ ] {n}{k}_{q_{i}}\,e_{i}^{n-k}q_{i}^{k%
\frac{h_{i}}{2}}\otimes e_{i}^{k}q_{i}^{-(n-k)\frac{h_{i}}{2}}  \label{uga}
\end{equation}
one verifies that the symmetry generators acting on the tensor product space 
$V^{\otimes L}$ are given by the expression 
\begin{equation}
E_{i}^{(N)}\equiv \Delta ^{(L)}(e_{i}^{(N)})=\sum_{0=n_{0}\leq n_{1}\ldots
\leq n_{L}=N}\tbigotimes_{l=1}^{L}e_{i}^{(n_{l}-n_{l-1})}q_{i}^{\left(
N-n_{l}-n_{l-1}\right) \frac{h_{i}}{2}}\;.  \label{sgen}
\end{equation}
This formula is immediate to derive by exploiting the fact that the
coproduct is an algebra homomorphism $U_{q}(\hat{g})\rightarrow U_{q}(\hat{g}%
)\otimes U_{q}(\hat{g})$ and then applying equation (\ref{uga}) to the first
factor in the tensor product. Note that according to the definition of $%
E_{i}^{(N)}$ we have divided out the factor $[N]_{i}!$ present in the $q$%
-binomial coefficient in (\ref{uga}). A similar formula holds for $%
F_{i}^{(N)}$.

\noindent We are now prepared to generalize the proof of invariance found in 
\cite{DFM}\ for the XXZ model in the fundamental representation. In order to
show translation invariance of the symmetry algebra it is obviously
sufficient to show that the generators $E_{i}^{(N^{\prime
})},F_{i}^{(N^{\prime })},H_{i}/N^{\prime }$ commute with the shift
operator. From (\ref{gen}) we see immediately that $[\Pi ,H_{i}]=0$ for all $%
q$. We now state the proof for $E_{i}^{(N^{\prime })}$, the one for $%
F_{i}^{(N^{\prime })}$ is completely analogous. For generic $q$ one finds
the following relations (compare \cite{DFM}) 
\begin{equation}
\Pi \,E_{i}\,\Pi
^{-1}=E_{i}\,q_{i}^{H_{i}(L)}+E_{i}(L)(q_{i}^{-H_{i}}-1)q_{i}^{H_{i}(L)}\;,
\end{equation}
where use has been made of the straightforward identities 
\begin{eqnarray*}
\Pi \,E_{i}(n)\,\Pi ^{-1} &=&E_{i}(n-1)q_{i}^{H_{i}(L)}\quad n>1 \\
\Pi \,q_{i}^{H_{i}(n)}\,\Pi ^{-1} &=&q_{i}^{H_{i}(n-1)}\;.
\end{eqnarray*}
We claim that the transformation property for the $m^{\text{th}}$ power of
the generator reads 
\begin{equation}
\Pi \,E_{i}^{m}\,\Pi
^{-1}=\sum_{n=0}^{m}E_{i}^{m-n}E_{i}(L)^{n}\,q_{i}^{n(m-1)}\QTATOPD[ ] {m}{n}%
_{q_{i}}q_{i}^{mH_{i}(L)}\prod_{l=0}^{n-1}\left( q_{i}^{-2l-H_{i}}-1\right)
\;.  \label{cshift}
\end{equation}
Here it is understood that the product yields one if $n=0$. For the proof we
proceed once more by induction. Assume that the above relation holds for $m$
we calculate 
\begin{multline*}
\Pi \,E_{i}^{m+1}\,\Pi
^{-1}=\tsum_{n=0}^{m}E_{i}^{m-n}E_{i}(L)^{n}E_{i}\,q_{i}^{n(m-1)}\QTATOPD[ 
] {m}{n}_{q_{i}}q_{i}^{(m+1)H_{i}(L)}\tprod_{l=1}^{n}\left(
q_{i}^{-2l-H_{i}}-1\right) \\
+\tsum_{n=0}^{m}E_{i}^{m-n}E_{i}(L)^{n+1}\,q_{i}^{n(m-1)}\QTATOPD[ ] {m}{n}%
_{q_{i}}\left( q_{i}^{2m-H_{i}}-1\right)
q_{i}^{(m+1)H_{i}(L)}\tprod_{l=1}^{n}\left( q_{i}^{-2l-H_{i}}-1\right)
\end{multline*}
Employing the commutation relations 
\begin{equation*}
E_{i}(L)^{n}E_{i}=q_{i}^{2n}E_{i}E_{i}(L)^{n}+E_{i}(L)^{n+1}(1-q_{i}^{2n})
\end{equation*}
and the elementary relation 
\begin{equation*}
\lbrack m+n]=[m]q^{-n}+[n]q^{m}
\end{equation*}
for $q$-deformed integers one derives the desired result (\ref{cshift}). Now
taking the limit $q^{N}\rightarrow 1$ one finds by setting $m=N^{\prime }$
from (\ref{cshift}) that 
\begin{equation}
\Pi \,E_{i}^{(N^{\prime })}\,\Pi ^{-1}=E_{i}^{(N^{\prime })}q_{i}^{N^{\prime
}H_{i}(L)}\;,  \label{Pinv}
\end{equation}
since the product always contains a vanishing factor due to the condition (%
\ref{con}). We discuss the effect of the factor $q_{i}^{N^{\prime }H_{i}(L)}$
for the cases of odd and even roots of unity separately.

\begin{description}
\item[$N$ odd. ]  For odd roots of unity, $N=N^{\prime }$, the factor is
always equal to one. Thus, we conclude that the constructed symmetry algebra 
$U(\hat{g})$ generated by the elements $\{E_{i}^{(N)},F_{i}^{(N)},H_{i}/N\}%
\cup \{1\}$ commutes with the shift operator. Recall that for this case the
symmetry algebra has been constructed in complete generality, i.e. for all
Kac-Moody algebras.

\item[$N$ even.]  For even roots of unity $q_{i}^{N^{\prime }H_{i}(L)}$
produces in general alternating signs. This can be seen e.g. from the
commutation relation (compare (Q1) in Section 2) 
\begin{equation}
E_{j}(L)(q_{i}^{N^{\prime }})^{H_{i}(L)}=(q_{i}^{N^{\prime
}})^{-A_{ij}}(q_{i}^{N^{\prime }})^{H_{i}(L)}E_{j}(L)\;.
\end{equation}
However, for the special case of the XXZ model $\hat{g}=A_{1}^{(1)}$ one has 
$|A_{ij}|=2$ for all $i,j$ from which we infer that the generators $%
E_{i}^{(N^{\prime })},F_{i}^{(N^{\prime })}$ of the symmetry algebra $%
U(A_{1}^{(1)})$ either commute or anticommute with the shift operator
depending on $V_{L}\cong \mathbb{C}^{n+1}$ being either of even or odd
highest weight $n\in \mathbb{N}$, respectively. Here $n=2s$ and $s$ is the
spin. This is accordance with the results obtained in \cite{DFM}, where the
fundamental representation $n=1$ has been considered.
\end{description}

\section{$U(\hat{g})$ symmetry at roots of unity}

We are now prepared to establish the $U(\hat{g})$ invariance of the
statistical model associated with the affine quantum group $U_{q}(\hat{g})$
as defined in (\ref{trans}). The crucial ingredient for this derivation is
the R-matrix which provides a solution of the Yang-Baxter equation (\ref{YB}%
). How every quantum group $U_{q}(\hat{g})$ gives rise to such a solution is
reviewed in the appendix in order to keep this article self-contained. The
proof of invariance then hinges on two observations, namely that the
generators of the symmetry algebra commute with the shift operator and the
quantum group invariance of the permuted R-matrix (see equation (\ref{qinv})
in the appendix.) This allows us to state the symmetry property for the
large class of solutions (e.g. \cite{slnR,Smat}) of the Yang-Baxter equation
(\ref{YB}) in a completely general fashion.

\subsection{The integrable model}

Suppose now we are given a trigonometric solution $R(u)$ of (\ref{YB})
associated with $U_{q}(\hat{g})$ and which acts on the tensor product $%
V\otimes V$ of some representation space $V$. We choose to normalize the $R$%
-matrix such that 
\begin{equation}
\lim_{u\rightarrow 0}R(u)=\pi \;,  \label{norm}
\end{equation}
where $\pi $ is the permutation operator. For affine quantum groups we
consider this regularity property always holds. The normalization (\ref{norm}%
) fixes the constant in (\ref{Ham}) to be zero. As is well known \cite
{Baxter,QISM}\ it follows directly from the Yang-Baxter equation that 
\begin{equation}
\lbrack T(u),T(v)]=0  \label{int}
\end{equation}
which implies the integrability of the model. The corresponding infinite set
of charges is defined by the following power series expansion of the
transfer matrix at vanishing rapidity $u=0$, 
\begin{equation}
\ln T(u)=\sum_{n=0}^{\infty }\frac{u^{n}}{n!}T^{(n)}\quad \text{with\quad }%
T^{(n)}=\left. \frac{\partial ^{n}}{\partial u^{n}}\ln T(u)\right| _{u=0}\;.
\label{exp}
\end{equation}

From equation (\ref{int}) we now immediately infer $%
[T(u),T^{(n)}]=[T^{(n)},T^{(m)}]=0$ for all choices of $n,m\in \mathbb{N}$
which manifests the integrability of the model. The zeroth and first order
term in the above expansion (\ref{exp}) are of special significance. From
the normalization condition (\ref{norm}) of the $R$-matrix one derives for
the zeroth term 
\begin{equation}
T^{(0)}=\ln T(0)=\ln \Pi ^{-1}\equiv -iP\;,  \label{mom}
\end{equation}
where $\Pi $ is the 'shift'-operator introduced in (\ref{shift}) and which
generates translations in the horizontal direction of the lattice. This
motivates the identification of $T^{(0)}$ as the momentum operator. The
first order term $T^{(1)}$ is identical with the (formal) spin-chain
Hamiltonian\footnote{%
Notice that this definition of the Hamiltonian is formal in the sense that
it is not necessarily always hermitian. For example, as was pointed out for
the XXZ model \cite{FYF} hermiticity might restrict for fixed spin the
allowed values of the coupling constant $\gamma $ incorporated in the
deformation parameter $q=e^{i\gamma }$.} as defined in the introduction (\ref
{Ham}), 
\begin{equation}
\mathcal{H}=i\left. \frac{\partial }{\partial u}\ln T(u)\right|
_{u=0}=i\sum_{j=1}^{L}\left. \frac{\partial }{\partial u}\mathcal{R}%
_{jj+1}(u)\right| _{u=0}\quad \text{with\quad }L+1\equiv 1\,.  \label{ham}
\end{equation}
Here we have defined the operator $\mathcal{R}_{jj+1}(u):=\pi
_{jj+1}R_{jj+1}(u)$ which acts on the tensor product $V_{j}\otimes V_{j+1}$
in the chain $V^{\otimes L}=V_{1}\otimes \cdots \otimes V_{L}$. Formula (\ref
{ham}) can be derived directly from the following operator identity over $%
V_{0}\otimes V_{1}\otimes \cdots \otimes V_{L}$%
\begin{equation}
R_{0L}(u)\cdots R_{01}(u)=\pi _{01}\Pi ^{-1}\,\mathcal{R}_{L-1L}(u)\cdots 
\mathcal{R}_{12}(u)\,\mathcal{R}_{01}(u)\;
\end{equation}
The main reason for changing from the original $R$-matrix to $\mathcal{R}%
(u)=\pi R(u)$ comes from the observation that the latter is invariant under
the quantum group action (see equation (\ref{qinv}) in the appendix). We
will now use this fact together with the notion of the boost operator to
demonstrate that all conserved charges of the model and especially the
transfer matrix are left invariant under the action of $U(\hat{g})$.

\subsection{The boost operator}

The boost operator of the integrable model is implicitly defined by the
relation 
\begin{equation}
-\frac{\partial }{\partial u}T(u)=[K,T(u)]\;.  \label{boost}
\end{equation}
Its explicit expression in terms of R-matrices has been found in e.g. \cite
{boost}, 
\begin{equation}
K=\sum_{n\in \mathbb{N}}\sum_{j=1}^{L}(j+nL)\partial _{u}\mathcal{R}%
_{jj+1}(u)|_{u=0}  \label{boost1}
\end{equation}
and can be derived by differentiating the Yang-Baxter equation (\ref{YB})
and exploiting translation invariance of the transfer matrix. The name
'boost' operator stems from the observation that $P,\mathcal{H},K$ form a
closed algebra which might be interpreted as a lattice version of the
Poincar\'{e} algebra (see e.g. the article by Thacker \cite{boost}). From
the defining property (\ref{boost}) one infers that under the adjoint action
of the boost operator the transfer matrix is shifted in the spectral
parameter, 
\begin{equation}
T(u+v)=e^{-vK}T(u)e^{vK}\;.  \label{boost2}
\end{equation}
Therefore, it is sufficient to show that the generators of the constructed
algebra $U(\hat{g})$ commute with the shift operator $\Pi =e^{iP}$ and the
boost operator $K$ in order to ensure that they also commute with the
transfer matrix and the Hamiltonian (as well as all higher charges). Since
we have already proven in Section 3 that for roots of unity with $N$ odd the
symmetry algebra is translation invariant, i.e. $[X,\Pi ]=0$ for all $X\in U(%
\hat{g})$, we need only to show the invariance of the boost operator.

\subsection{Invariance of the boost operator}

According to equation (\ref{qinv}) in the appendix one easily verifies that
the operators $\mathcal{R}_{jj+1}(u)$ commute with the generators (\ref{gen}%
) of the quantum group $U_{q}(\hat{g})^{\otimes L}$ in the evaluation
representation at generic $q$ for $j<L$, 
\begin{equation}
\lbrack \Delta _{u}^{(L)}(x),\mathcal{R}_{jj+1}(u)]=0\;,\quad x\in U_{q}(%
\hat{g}),\;j<L  \label{QGinv}
\end{equation}
Recall that when taking the root of unity limit a universal R-matrix may not
always exist, since then the elements $E_{i}^{N},F_{i}^{N},q_{i}^{\pm H_{i}}$
are central and the quasitriangular structure imposes constraints on their
spectral values, see e.g. the article by E. Date et al. \cite{Smat}.
However, we are only interested in representations where $%
E_{i}^{N},F_{i}^{N}\equiv 0$. Then these restrictions do not exist and a
universal R-matrix can be defined. Therefore, property (\ref{QGinv}) remains
true when the limit $q^{N}\rightarrow 1$ is taken and, consequently, all we
need to show is that $\mathcal{R}_{L1}(u)$ commutes with $X\in U(\hat{g})$.
But since we have established translation invariance of the symmetry algebra
in the root of unity limit, one immediately verifies that 
\begin{eqnarray}
\lbrack X,\mathcal{R}_{L1}(u)] &=&\Pi ^{-1}[X,\Pi \,\mathcal{R}_{L1}(u)\,\Pi
^{-1}]\,\Pi  \notag \\
&=&\Pi ^{-1}[X,\mathcal{R}_{L-1L}(u)]\,\Pi =0\;,\quad X\in U(\hat{g})
\label{Qinv2}
\end{eqnarray}

From (\ref{boost1}) and the fact that $\mathcal{R}(0)=1$ we conclude that
the boost operator commutes with all elements of the algebra $U(\hat{g})$ in
the evaluation representation with $e^{u}\rightarrow 1$. Notice that by the
same arguments it also follows directly that the Hamiltonian (\ref{ham}) is
invariant. We thus conclude that the integrable model (\ref{trans})
associated with the quantum group $U_{q}(\hat{g})$ exhibits an $U(\hat{g})$
invariance as $q^{N}\rightarrow 1$ with $N$ being odd.

\subsection{Even roots of unity and $\hat{g}=A_{1}^{(1)}$}

In Section 2 we have seen that the construction of the algebra $U(\hat{g})$
for the XXZ model $\hat{g}=A_{1}^{(1)}$ also holds for even roots of unity.
As has beome apparent in Section 3 the difference to case of odd roots
occurs when the behaviour of the symmetry algebra under translation is
investigated. Depending on the choice of the spin $s=n/2$ the constructed
algebra generators $E_{i}^{(N^{\prime })},F_{i}^{(N^{\prime })},\,i=0,1$
commute or anticommute with the shift operator if we choose the
representation spaces $V_{n}$ in the $L$-fold tensor product to be of
highest weight $n$ even or odd. It is evident from equation (\ref{boost2})
and (\ref{Qinv2}) that in the latter case the symmetry algebra still
commutes with the boost operator, but that the generators $E_{i}^{(N^{\prime
})},F_{i}^{(N^{\prime })}$ now anticommute with the transfer matrix. We
therefore conclude that 
\begin{equation}
XT(u)=(-1)^{2s}T(u)X\;,\quad X=E_{i}^{(N^{\prime })},F_{i}^{(N^{\prime })}
\end{equation}
which generalizes the results obtained in \cite{DFM} to arbitrary spin $s\in 
\frac{1}{2}\mathbb{N}$. In contrast, the Hamiltonian (\ref{ham}) obviously
commutes with all elements of the symmetry algebra for even and odd roots
independent of the chosen spin value.

\section{Conclusions}

In this article we have shown that the loop symmetry of the Hamiltonian and
the transfer matrix first observed in the context of the XXZ and the six
vertex model for spin $s=1/2$ at roots of unity \cite{DFM} is of a general
nature. For odd roots of unity we demonstrated that it is present for
generic integrable vertex models associated with trigonometric solutions of
the Yang-Baxter equation with underlying quantum group $U_{q}(\hat{g})$, $%
\hat{g}$ being any Kac-Moody algebra. The invariance has been shown to be a
direct consequence of both the quasi-triangular structure of the quantum
group $U_{q}(\hat{g})$ and the translation invariance of the symmetry
algebra $U(\hat{g})$. While for generic algebras we had to restrict
ourselves for the construction to roots of unity $q^{N}=1$ with $N$ odd the
loop symmetry could be extended also to $N$ even for the XXZ model thereby
generalizing the results obtained for the fundamental representation \cite
{DFM} to arbitrary spin. The restriction on the highest weight
representations (\ref{con}) might only be of technical nature since the
numerical investigations performed in the context of the XXZ model \cite
{DFM,FM} point out that the loop symmetry is present in general. We expect,
however, the construction of the symmetry algebra in the other cases to be
more involved.

We therefore conclude that the spectrum of the Hamiltonian (\ref{Ham})
organizes in multiplets of finite dimensional representations of $U(\hat{g})$%
. That is, given an eigenstate of the Hamiltonian or transfer matrix it
belongs to some highest weight representation $\lambda =\left( \lambda
_{1},...,\lambda _{r}\right) $ with $r=\limfunc{rank}g$ being the rank and $%
\lambda _{i}=\lambda (H_{i})$ the eigenvalues of the Cartan generators $%
H_{i} $ acting on $V^{\otimes L}$. (We assume that $\lambda $ satisfies the
condition $\lambda _{i}\func{mod}N=0$.) The irreducible finite dimensional
representations of untwisted algebras have been shown to be isomorphic to
tensor products of evaluation representations 
\begin{equation}
V_{\Lambda _{1}}(a_{1})\otimes V_{\Lambda _{2}}(a_{2})\cdots \otimes
V_{\Lambda _{l}}(a_{l})\;,
\end{equation}
whose evaluation parameters $a_{k},\,k=1,...,l$\ are determined by the roots
of $\limfunc{rank}g$ Drinfel'd Polynomials (see \cite{CP2}, pp 11). Here $g$
is the finite dimensional algebra whose affinization is $\hat{g}$ and $l$ is
some integer depending on the multiplet $\lambda $ of the energy eigenstate.
Let $n_{i,k}$ be the multiplicity of the root $a_{k}$ in the $i^{\text{th}}$
Polynomial then the weights appearing in the above product are given by $%
\Lambda _{k}=\sum_{i=1}^{\limfunc{rank}g}n_{i,k}\omega _{i}$ with $\omega
_{i}$ denoting the fundamental weights of $g$. Thus, the dimension of the
representation, i.e. the degeneracy of the energy eigenstate is given by 
\begin{equation}
\prod_{k=1}^{l}\dim \Lambda _{k}\quad \text{with\quad }\dim \Lambda
=\prod_{\alpha >0}\frac{\left\langle \Lambda +\varrho ,\alpha \right\rangle 
}{\left\langle \varrho ,\alpha \right\rangle }\;.
\end{equation}
Here the dimension is calculated from the Weyl formula with the product
running over the positive roots of $g$ and $\varrho $ denotes the Weyl
vector. The numerical work in the context of the XXZ model \cite{DFM,FM}
indicates that in the above tensor product only the fundamental
representation appears, explaining the degeneracy factors $2^{l}$.

Analogous to the work for the spin $s=1/2$ XXZ model one can now investigate
the relation of the generally established symmetry to the Bethe Ansatz. As
mentioned already in the introduction it was found \cite{DFM,FM} that the
degeneracies are related to ambiguities inside Bethe's equation which fails
to determine the complete set of string solutions at roots of unity. The
missing solutions were shown to be complete exact $N$-strings 
\begin{equation*}
v_{k}^{(N)}=\alpha +ik\,2\gamma \;,\quad 1\leq k\leq N,\,q=e^{i\gamma },\,%
\frac{\gamma }{\pi }=\frac{m}{N}\in \mathbb{Q}
\end{equation*}
which have momentum $P=0,\pi $ and zero energy and give thus rise to
degeneracies in the energy spectrum. These observations can also be
generalized. For example we observe that Bethe's equation for the higher
spin XXZ model \cite{KiResh} 
\begin{equation*}
\left( \frac{\sinh \frac{1}{2}\left( v_{j}+i2s\gamma \right) }{\sinh \frac{1%
}{2}\left( v_{j}-i2s\gamma \right) }\right) ^{L}=\prod_{k\neq j}^{sL-|S^{z}|}%
\frac{\sinh \frac{1}{2}\left( v_{j}-v_{k}+i2\gamma \right) }{\sinh \frac{1}{2%
}\left( v_{j}-v_{k}-i2\gamma \right) }
\end{equation*}
incorporates the spin dependence only on the l.h.s. while the ambiguous
factors $0/0$ due to the complete exact $N$-strings occur on the r.h.s. of
the above equation. Hence, the statements made in the context of the spin $%
s=1/2$ case can be generalized in a straightforward way. In particular, the
derivation of the equations determining the real parts $\alpha $ of the
complete exact $N$-strings in the limit $q^{N}\rightarrow 1$ follows exactly
along the lines of \cite{FM}. In view of these considerations one might now
anticipate that the regular solutions to the Bethe's equation give the
highest weight states w.r.t. to the symmetry algebra and that the action of
the latter corresponds to adding complete exact $N$-strings to this
solution. However, to understand this relation fully in terms of
representation theory of the symmetry algebra and how to extract from the
Bethe Ansatz the evaluation parameters $a_{k}$ are both open problems at the
moment. They require a deeper and more profound understanding of the quantum
group theoretical structure of the Bethe Ansatz.\bigskip

\noindent \textbf{Acknowledgments}: The first author (C.K.) would like to
thank I. Roditi for useful discussions and J. Links for helpful comments on
the boost operator. The second author (B.M.M.) is deeply indebted to K.
Fabricius and T. Deguchi for many insights into symmetries of lattice models
and to M. Jimbo, M. Kashiwara, and T. Miwa for fruitful discussions on
quantum groups. He also wishes to acknowledge the hospitality of the
Research Institute of Mathematical Sciences of Kyoto University where this
work was completed. The financial support of the National Science Foundation
(Grants DMR-0073058 and PHY-9988566) is gratefully acknowledged.

\appendix

\section{The Yang-Baxter equation}

In order to keep this paper self-contained we briefly review the connection
between quantum groups and trigonometric solutions to the Yang-Baxter
equation {\cite{Drin,Jimbo}}.

First one introduces for arbitrary spectral parameter $u$ the following
automorphism $D_{u}$ on $U_{q}(\hat{g})$ which equals the identity on all
generators except for 
\begin{equation}
D_{u}(e_{0})=e^{u}\,e_{0}\quad \text{and}\quad
D_{u}(f_{0})=e^{-u}\,f_{0}\quad .
\end{equation}
Loosely, speaking it can be thought of as conjugating by $\left(
e^{u}\right) ^{d}$, where $d$ is the homogeneous degree operator of the
affine Lie algebra $\hat{g}$. Analogously, one might also consider the
principal grading $\hat{\varrho}=h^{\vee }d+\varrho $ where $h^{\vee }$ is
the dual Coxeter number and $\varrho $ the Weyl vector, see e.g. \cite{MJ}.
Suppose we are given a finite-dimensional representation $\rho _{V}:U_{q}(%
\hat{g})\rightarrow \limfunc{End}(V)$, such that $\rho _{V}$ viewed as
representation of $U_{q}(g)$ has finite length and all irreducible
subrepresentations are highest weight \cite{EFK}. Here $g$ denotes the
finite dimensional simple Lie algebra whose affinization is $\hat{g}$. Then
an evaluation representation $V(u)$ is defined through the following
composition of maps, 
\begin{equation}
\rho _{V(u)}=\rho _{V}\circ D_{u}\;.
\end{equation}
Originally the finite-dimensional representations $\rho _{V}$ have been
explicitly constructed via evaluation homomorphisms $p:U_{q}(\hat{g}%
)\rightarrow U_{q}(g)$ by Jimbo \cite{Jimbo} for the series $\hat{g}%
=A_{n}^{(1)}$. Since they have been implicitly used for the construction in 
\cite{DFM}, we review them at the end of this section. However, these
evaluation homomorphisms do not exist in general \cite{CP,EFK} whence the
above definition of evaluation representations is the more generic one.

Secondly we must specify how solutions of the Yang-Baxter equation come
about in the setting of evaluation representations. The R-matrix naturally
arises in the context of the coproduct structure (\ref{copo}) which is part
of the definition of $U_{q}(\hat{g})$ as Hopf algebra \cite{Drin,Jimbo}. As
already mentioned in the main text the latter is needed to build up tensor
products of representations giving the state space $V^{\otimes L}$ of the
statistical model. As we infer from the definition of the coproduct (\ref
{copo}) these tensor products carry an 'orientation' since the quantum group 
$U_{q}(\hat{g})$ viewed as an Hopf algebra is in general non-cocommutative.
In formulas, this means that the action of the \textbf{'}opposite\textbf{' }%
coproduct 
\begin{equation}
\Delta ^{\text{op}}\equiv \pi \circ \Delta ,  \label{op}
\end{equation}
where $\pi $ denotes the permutation operator, does not coincide with the
action of $\Delta $. Thus, the products $V\otimes W$ and $W\otimes V$ of two
representations $V,W$ are distinct. However, as is well known the quantum
groups belong to the class of quasi-triangular Hopf algebras where the two
different coproduct structures can be related by conjugation via some
invertible element, the 'universal $R$-matrix' $R\in U_{q}(\hat{g})\otimes
U_{q}(\hat{g})\;$(see e.g. \cite{CP} for further details), 
\begin{equation*}
\Delta ^{\text{op}}(x)=R\,\Delta (x)\,R^{-1}
\end{equation*}
This universal $R$-matrix as well as the coproduct acquire a spectral
parameter dependence when we specialize to evaluation representations $%
V(u),W(v)$. We emphasize that both $V$ and $W$ are considered to be finite
dimensional, whence we work in a level zero representation of $U_{q}(\hat{g}%
) $. The spectral parameter dependent R-matrix and coproduct are then
defined via 
\begin{equation}
R_{V,W}^{{}}(u-v):=(\rho _{V(u)}\otimes \rho _{W(v)})R\quad \text{and\quad }%
\Delta _{V(u),W(v)}=(\rho _{V(u)}\otimes \rho _{W(v)})\Delta  \label{zdef}
\end{equation}
The intertwining property for the coproduct $\Delta $ and its counterpart $%
\Delta ^{\text{op}}$ then reads 
\begin{equation}
R_{V,W}^{{}}(u-v)\Delta _{V(u),W(v)}(x)=\Delta _{V(u),W(v)}^{\text{op}%
}(x)R_{V,W}^{{}}(u-v)\;,\quad x\in U_{q}(\hat{g})\,.  \label{Rinter}
\end{equation}
As indicated the R-matrix depends only on the difference $u-v$ (see e.g. 
\cite{Jimbo}), whence we might set $v=0$ in the above relations. In
addition, we choose in what follows $V=W$ and drop the dependence on the
representation $V$ in order to unburden the notation. Besides the
intertwining property (\ref{Rinter}) the $R$-matrix is subject to several
other requirements. For example, in order that the quasi-triangular
structure is compatible with coassociativity, $\left( \Delta \otimes \mathbf{%
1}\right) \Delta =\left( \mathbf{1}\otimes \Delta \right) \Delta $, the $R$%
-matrix is subject to the following equations sometimes referred to as
'fusion laws' (see e.g. \cite{CP,EFK}), 
\begin{eqnarray}
(\Delta _{u}\otimes \mathbf{1})(R(v)) &=&R_{13}(u+v)R_{23}(v)  \notag \\
(\mathbf{1}\otimes \Delta _{u})(R(v)) &=&R_{13}(v-u)R_{12}(v)\;,\quad
\label{R2}
\end{eqnarray}
The Yang-Baxter equation (\ref{YB}) is now an immediate consequence of the
first identity and the intertwining property. This establishes the link
between affine quantum groups and integrable models as found by Drinfel'd
and Jimbo.

We conclude by recalling that from the defining relations (\ref{op}), (\ref
{zdef}), and the property (\ref{Rinter}) it is straightforward to check that
for $V=W$ the composition $\mathcal{R}(u):=\pi R(u)$ is quantum group
invariant \cite{Jimbo} 
\begin{equation}
\lbrack \mathcal{R}(u),\Delta _{u}(x)]=0\;,\quad x\in U_{q}(\hat{g})\;.
\label{qinv}
\end{equation}
Here $\pi $ is the permutation operator introduced earlier. It is this
invariance property of the intertwiner $\mathcal{R}(u)$ under $U_{q}(\hat{g}%
) $ which we exploit to prove the invariance of the transfer matrix and its
associated higher charges under the algebra $U(\hat{g})$ at roots of unity.
We also note that if $(U_{q}(\hat{g}),\Delta _{u},R(u))$ forms a
quasitriangular Hopf algebra, so do the combinations $(U_{q}(\hat{g}),\Delta
_{u}^{\text{op}},R(u)^{-1})$ and $(U_{q}(\hat{g}),\Delta _{u}^{\text{op}},R^{%
\text{op}}(u))$ where $R^{\text{op}}(u)=\pi R(u)\pi $, see e.g. \cite{CP}.
Hence, one has the additional relations $[\pi \,R(u)^{-1},\Delta _{u}^{\text{%
op}}(x)]=[\pi R^{\text{op}}(u),\Delta _{u}^{\text{op}}(x)]=0$.

\subsection{The evaluation representation for $U_{q}(A_{1}^{(1)})$}

As a concrete example for the abstract considerations outlined above we now
explicitly state the evaluation representation for the quantum group $%
U_{q}(A_{1}^{(1)})$. We will start from the evaluation representation $%
p_{z}:U_{q}(A_{1}^{(1)})\rightarrow U_{q}(A_{1})$ found by Jimbo \cite{Jimbo}
and then construct in a second step an evaluation representation for the $L$%
-fold tensor product $p_{z}^{(L)}:U_{q}(A_{1}^{(1)})^{\otimes L}\rightarrow
U_{q}(A_{1})^{\otimes L}$ which is the physical case of interest since the
transfer matrix lives as an operator on the quantum state space $V^{\otimes
L}$. Therefore, we need to construct a representation of our symmetry
algebra which acts on the same space. Note that this is not simply
accomplished by taking the $L$-fold tensor product of $p_{z}$, since the
evaluation homomorphism does not constitute an Hopf algebra homomorphism,
i.e. the coproduct structure is not preserved \cite{Jimbo}.

The quantum group $U_{q}(A_{1}^{(1)})$ consists of all power series in terms
of the generators $\{e_{i},f_{i},h_{i}\}_{i=0,1}\cup \{1\}$. The commutation
and Chevalley-Serre relations can be deduced from the Cartan matrix $%
A=\left( 
\begin{array}{cc}
2 & -2 \\ 
-2 & 2
\end{array}
\right) $ according to the definitions (Q1)-(Q3). The homomorphism found in 
\cite{Jimbo} reads 
\begin{eqnarray}
e_{0} &\rightarrow &p_{u}(e_{0})=e^{u}\,f\quad f_{0}\rightarrow
p_{u}(f_{0})=e^{-u}\,e\quad h_{0}\rightarrow p_{u}(h_{0})=-h\;,  \notag \\
e_{1} &\rightarrow &p_{u}(e_{1})=e\quad f_{1}\rightarrow p_{u}(f_{1})=f\quad
h_{1}\rightarrow p_{u}(h_{1})=h
\end{eqnarray}
It is a straightforward exercise to verify that the above identification of
the generators not only preserve the commutation relations of $%
U_{q}(A_{1}^{(1)})$ but also the Chevalley-Serre relations. For simplicity
we set in the following $u=0$, since this is the only case which will be
relevant later on. The generalization to non-zero spectral parameter is
straightforward.

To construct now the generators of the $L$-fold tensor product $%
U_{q}(A_{1}^{(1)})^{\otimes L}$ we apply the coproduct $\Delta $ iteratively
as defined in (\ref{copL}) and obtain the generators (\ref{gen}) for $i=0,1$%
. Remembering that we have $h_{0}=-h_{1},e_{0}=f_{1}$ and $f_{0}=e_{1}$ in
the evaluation representation we might rewrite the generators (\ref{gen}) in
terms of the generators of $U_{q}(A_{1})$ by exploiting both of the
non-affine coproduct structures $\Delta ^{\prime }$ and $\Delta ^{\prime 
\text{op}}=\pi \circ \Delta ^{\prime }$. The latter are obtained by
restricting the affine coproduct $\Delta $ to the subalgebra $%
\{e_{1}=e,f_{1}=f,h_{1}=h\}$. The following identities then hold 
\begin{eqnarray}
E_{0} &\equiv &\Delta ^{\prime \text{op}(L)}(f)\;,\quad F_{0}\equiv \Delta
^{\prime \text{op}(L)}(e)\;,\quad \text{and\quad }q^{H_{0}}\equiv \Delta
^{\prime (L)}(q^{-h})  \notag \\
E_{1} &\equiv &\Delta ^{\prime (L)}(e)\;,\quad F_{1}\equiv \Delta ^{\prime
(L)}(f)\;,\quad \text{and\quad }q^{H_{1}}\equiv \Delta ^{\prime (L)}(q^{h})
\end{eqnarray}
Here we have made use of the fact that the opposite coproduct $\Delta
^{\prime \text{op}}=\pi \circ \Delta ^{\prime }$ of $U_{q}(A_{1})$ is easily
seen to be obtained by formally replacing $q\rightarrow q^{-1}$ (compare (%
\ref{copo})). This establishes an homomorphism $U_{q}(A_{1}^{(1)})^{\otimes
L}\rightarrow U_{q}(A_{1})^{\otimes L}$, since $\Delta ^{\prime (L)},\Delta
^{\prime \text{op}(L)},\Delta ^{(L)}$ are all algebra homomorphisms. In
particular, the affine quantum Chevalley-Serre relations of $%
U_{q}(A_{1}^{(1)})^{\otimes L}$ are also satisfied in $U_{q}(A_{1})^{\otimes
L}$ under this representation. This matches the construction of $%
U_{q}(A_{1}^{(1)})^{\otimes L}$ obtained in \cite{DFM}\ for the fundamental
representation $V\cong \mathbb{C}^{2}$.

\subsection{The evaluation representation for $U_{q}(A_{n}^{(1)})$}

Following Jimbo \cite{Jimbo} we recall for completeness how the evaluation
representations for the higher rank algebras are constructed. The
homomorphism $p:U_{q}(A_{n}^{(1)})\rightarrow U_{q}(A_{n})$ is constructed
by defining iteratively elements $\frak{E}_{ij}\in U_{q}(A_{n})$, $i\neq j$:
Setting $\frak{E}_{ii+1}:=e_{i}$ and $\frak{E}_{i+1i}:=f_{i}$ assume that
these elements had been constructed for $|i-j|<k$, then the elements with $%
|i-j|=k$ are given by 
\begin{equation}
\frak{E}_{ij}=\left\{ 
\begin{array}{cc}
\frak{E}_{ii+1}\frak{E}_{i+1j}-q\frak{E}_{i+1j}\frak{E}_{ii+1}\;, & i<j \\ 
\frak{E}_{ii-1}\frak{E}_{i-1j}-q^{-1}\frak{E}_{i-1j}\frak{E}_{ii-1}\;, & i>j
\end{array}
\right. \;.
\end{equation}
It has been shown \cite{Jimbo} that the so defined elements satisfy the
relation 
\begin{equation}
\frak{E}_{ij}=\frak{E}_{il}\frak{E}_{lj}-q^{\pm 1}\frak{E}_{lj}\frak{E}%
_{il}\;,\quad i\lessgtr l\lessgtr j
\end{equation}
from which one may verify that the mapping 
\begin{equation}
p(e_{0})=\frak{E}_{n1}\;,\quad p(f_{0})=\frak{E}_{1n}\quad \text{and\quad }%
p(q^{h_{0}})=\prod_{j=1}^{n-1}q^{-h_{j}}
\end{equation}
fixes an algebra homomorphism $U_{q}(A_{n}^{(1)})\rightarrow U_{q}(A_{n})$
when identifying the remaining generators in the natural way, $%
p(e_{i})=e_{i},p(f_{i})=f_{i},p(q^{h_{i}})=q^{h_{i}}$ with $i>0$. One might
then apply the affine coproduct (\ref{copo}) in this evaluation
representation in order to obtain expressions for the generators in the
physical state space analogous to (\ref{gen}) and (\ref{sgen}). Note that
the interplay between the affine coproduct $\Delta $ and the non-affine
coproducts $\Delta ^{\prime },\Delta ^{\prime \text{op}}$ is special to the $%
A_{1}\equiv sl_{2}$ case and does not apply for $n>1$.

\end{document}